\def \CMP {{\it Commun. Math. Phys.}} 
\def \JMP {{\it J. Math. Phys. }}

\def \bc {\begin{center}} 
\def \ec {\end{center}}

\def \bfr {\begin{flushright}}
\def \efr {\end{flushright}}

\def \v {\vskip}

\def \ba {\begin{array}}
\def \ea {\end{array}}

\def \bea {\begin{eqnarray}}
\def \eea {\end{eqnarray}}

\def \be {\begin{equation}}
\def \ee {\end{equation}}
\def \p {\partial}

\def \L {{\cal L}}
\def \d {\hbox{d}\,}

\def \e {\hbox{e}}

\def \M {{\cal M}}

\def \tG {\widetilde{G}}

\def \F {{\cal F}}

\def \T {{\cal T}}
\def \H {{\cal H}}

\def \Diff {{\hbox{Diff}}}
 
\newcommand{\bb}[1]{{\bf \bar{#1}}}
\documentstyle[12pt,a4]{article}
\begin{document}

\catcode`\@=11
\@addtoreset{equation}{section}
\def\theequation{\arabic{section}.\arabic{equation}}
\catcode`\@=11

\def \thesection {\Roman{section}} 
\def \thesubsection {\Alph{subsection}}
\def \thesubsubsection {\thesubsection.\arabic{subsubsection}} 

\newcounter{definition}[section]
\def \Def {\stepcounter{definition} 
\noindent {\it Definition} \arabic{section}.\arabic{definition}} 

\newcounter{theorem}[section]
\def \Theorem {\stepcounter{theorem} 
\noindent 
{\it Theorem} \arabic{section}.\arabic{theorem}}

\newcounter{proposition}[section]
\def \Prop {\stepcounter{proposition} \noindent 
{\it Proposition} \arabic{section}.\arabic{proposition}}

\newcounter{corollary}[theorem]
\def \Corollary {\stepcounter{corollary} \noindent 
{\it Corollary} \arabic{section}.\arabic{theorem}.\arabic{corollary}}

\newcounter{lemma}[section]
\def \Lemma {\stepcounter{lemma} \noindent 
{\it Lemma} \arabic{section}.\arabic{lemma}}

\def \Proof {\noindent{\it Proof}}
\def \Note {\noindent{\it Note}}

\thispagestyle{empty}
\hfil{{\bf Imperial-TP/95-96/12}}\break 
\vskip15mm
\begin{center}

\Large 

{\bf Group Quantization on Configuration Space:} 
\v2mm
{\bf Gauge Symmetries and Linear Fields\footnote[1]{Work partially 
supported by the Comisi\'on
Interministerial de Ciencia y Tecnolog\'\i a.}}
\normalsize
\vskip7mm
{\it 
Miguel Navarro\footnote{http://www.ugr.es/$\widetilde{\>}$mnavarro; 
e-mail: m.navarro@ugr.es}$^{1,2,3}$,  
V\'\i ctor Aldaya\footnote{e-mail: valdaya@ugr.es}$^{3,4}$ 
and Manuel Calixto\footnote{e-mail: calixto@ugr.es}$^{3,5}$ 
}  
\vskip5mm

\end{center}

\begin{enumerate}

\item The Blackett Laboratory, Imperial College, Prince 
Consort Road, London SW7 2BZ; United Kingdom.

\item Instituto de Matem\'aticas y F\'\i sica Fundamental, CSIC, Serrano
    113-123, 28006 Madrid, Spain

\item Instituto Carlos I de F\'\i sica Te\'orica y Computacional,
Facultad  de  Ciencias, Universidad de Granada, Campus de Fuentenueva,
18002, Granada, Spain.

\item IFIC, Centro Mixto Universidad de
Valencia-CSIC, Burjassot 46100-Valencia, Spain.

\item Departamento de F\'\i sica Te\'orica y del Cosmos, Facultad
de Ciencias, Universidad de Granada, 
Campus de Fuentenueva, Granada 18002, Spain

\end{enumerate}
\newpage
\thispagestyle{empty}
\centerline{}

\vskip10mm
\centerline{\bf Abstract}
A new, configuration-space picture of a formalism of 
group quantization, the GAQ formalism, is presented in the context of  
a previous, algebraic generalization. This presentation serves to make a
comprehensive discussion in which other 
extensions of the formalism, principally to incorporate  
gauge symmetries, are developed 
as well. Both images are combined in order to analyse, 
in a systematic manner and with complete generality, 
the case of linear fields (abelian current groups). To ilustrate these 
developments we particularize them for several fields and, in
particular, we carry out
the quantization of the abelian Chern-Simons models over an
arbitrary closed surface in detail. 
\v3mm
\noindent PACS numbers: 11.15.-q, 03.70.+k, 02.20.Tw

\normalsize
\vfil\eject 
\setcounter{page}{1}

\section{Introduction}
At present the main goal of 
Theoretical Physics is to unify Quantum 
Theory and General Relativity. Symmetry is increasingly important  
in both theories and, because of that, it is expected to 
play a principal role in the future fundamental theory whatever it  
might be. Therefore it is desirable to 
understand as much as possible about Physics without using information 
other than that provided by the symmetries of the systems. 
The formalisms of quantization on a group, such as the Group 
Approach to Quantization (GAQ) formalism, are intended to perfom 
this task as far as the process of quantization is concerned. 

The GAQ formalism was introduced several years
ago \cite{[JMP]} as an improved
version, in some respects, of Geometric Quantization and the Kirillov
coadjoint-orbit methods of quantization 
\cite{[Woodhouse],[Kirillov]}.  
It is conceived basically as an algorithm 
for associating quantum systems with already given groups.   
However, most classical systems are commonly 
specified by a set of different
equations or by a classical Lagrangian. 
Therefore in order to quantize these system with 
the GAQ formalism, it would be important to be able 
to derive, from the equations of motion or the 
Lagrangian, a group naturally associated with the
system and large enough so as to reproduce, in some way, the classical
theory.
 In so doing, solving the classical equations of 
motion has been required up to now.  Nevertheless, 
in ref. \cite{[config]} indications have been presented   
that there must be a way of circumventing 
this difficulty so that the basic steps, at least, of  
the GAQ formalism -- such as finding out the quantizing group --  
may be carried out without previously solving the equations 
of motion. This procedure constitutes the
{\bf configuration-space picture} of the GAQ formalism  
and its further development is the main purpose of the present  
paper. As our first step,  we shall 
consider linear fields only while   
non-abelian fields will be analysed in future studies. 

An improvement of the GAQ formalism which is specially relevant to our 
purposes is its algebraic reformulation, which, instead of the  
infinitesimal calculus, uses the finite (algebraic) properties 
of the group \cite{[CMP]}. 
This reformulation, therefore, enables us to
incorporate discrete symmetries and to deal with 
non-Lie groups, that is, groups with no differential structure. 
The basic aspects of this reformulation 
were previously presented in ref. \cite{[CMP]}. 
Here this picture of the formalism 
is presented in a unified manner so as to clarify several 
previous, heterogeneous developments. 
To make the discussion as self-contained as possible, 
the algebraic formulation is also further developed,    
in particular the characterization 
of gauge symmetries (gauge subgroup) is presented,  
and the way in which the GAQ formalism incorporates  
them at the quantum level is also shown. 

When working in configuration space, with no explicit 
expression for the group in terms of the phase-space coordinates of the 
fields, to use the 
differential calculus over this group is clearly not  
feasible. It is necessary, therefore, to use algebraic group 
transformations. This fact provides additional support for using the 
algebraic picture of the GAQ formalism. 

The quantization of linear fields, unlike non-linear ones 
whose quantization is considered to be a completely different 
and a much more difficult problem, 
is generally assumed to be well understood. 
There is in fact one good reason for such a
different behaviour between one case and the other: 
the huge (abelian) symmetry which underlies abelian fields. 
However, in spite of this fact, 
the usual way of presenting the quantization of linear
fields does not make it explicit whether or not 
this underlying symmetry is involved. 
This fact does not help to identify the real difficulties 
in quantizing non-linear fields. 
Also, if the difference lies in  
the great symmetry which underlies linear fields, 
we should examine whether or not it is possible
to construct non-linear fields, related to non-abelian current groups, 
which could be quantized with procedures similar to those applied in the
linear case. 

In addition to all this, and in spite of the (almost) general assumption, 
the quantization of linear fields is not always so trivial. 
There are many important cases, such as the one of 
fields in curved space (see for instance \cite{[Wald2]}), 
or when topological issues arise, in which the 
quantization presents difficulties 
and ambiguities with no simple solution.  

The motivation to study linear fields is therefore twofold: on the one
hand, they are important on their own, and,  
on the other hand this analysis 
may provide the key to generalize to non-linear fields. 

In the present paper linear fields are thoroughly studied,  
relying as much as possible 
on their underlying symmetry  and 
trying to be as general as possible. 
The structure of this paper is as follows: 
In Part 1, after a
brief review of the Geometric Quantization and the GAQ formalism
over a connected Lie group, the algebraic and configuration-space  
pictures of the GAQ formalism are considered. 
The results of this part are valid for arbitrary groups and fields. 
In Part 2, the theory of linear fields is thoroughly analysed by applying 
to it the (algebraic) GAQ formalism on configuration space. 
As an ilustration of how to apply the formalism,  
several aspects of the electromagnetic field 
are briefly considered in section \ref{Maxwell} -- 
the interested reader may also consult ref. \cite{[fieldsJPA]} and, above
all, refs. {\cite{[empro],[BRST]} where the development in this section
have been carried further --  
and the Abelian Chern-Simons theory 
is quantized in section \ref{Chern-Simons}. 
For the sake of clarity, in this part --except
in the last section-- the analysis is restricted to linear (abelian)
fields, even though one of our main 
motives is to extend, in the future, as much as possible of our results  
to non-abelian fields.   
In the last section, we discuss 
very briefly the difficulties in trying to extend our 
formalism to non-linear fields (non-abelian current groups).  

Since this paper is aimed to present the unifying theory behind    
some previous or parallel (and to clear the way to future) 
developments of the GAQ formalism -- those which only involve   
linear fields -- the examples has been 
carried on only up to the point that they provide a link with those 
developments but do not significantly overlap with them.  
For more details on how the GAQ formalism is actually applied,  
the reader may consult the bibliography 
here provided where diverse applications can be found. 
In particular ref. \cite{[config]},  where quite a few 
examples of quantizing groups in configuration space 
are also given, complements the present analysis in several
respects. 

\newpage
\centerline{}
\v5mm
\bc
\large {\bf PART 1. THE GENERAL FORMALISM} \normalsize\ec

\section{The Geometric Quantization and the 
Group Approach to Quantization} 
Before considering the GAQ formalism, 
we shall briefly describe 
the basic features of Geometric Quantization (GQ) which is a 
formalism from which the former derived. 

\subsection{Geometric Quantization}
The Geometric Quantization (see for instance
\cite{[Woodhouse]}) is a formalism which intends to place 
the familiar canonical quantization rules of Quantum Mechanics 
in a rigorous setting:

\bea q^i\longrightarrow {\widehat{q}}^i;&\quad& 
\left({\widehat{q}}^i \Psi\right)({\bf q}) \equiv q^i\Psi({\bf
q})\nonumber\\
p_j\longrightarrow\widehat{p}_j;&\quad&\left(\widehat{p}_j 
\Psi\right)({\bf q}) \equiv 
-i\hbar\frac{\p}{\p q^j}\Psi({\bf q})\label{rulesofq}\eea
where $q^i, p_j$ fulfil the classical relationships   
\be \{p_i,\>q^j\}=\delta_i^j\label{comcanonical}\ee 
[From here on we shall make $\hbar=1$.] 

The basic idea in this formalism is that the quantum 
theory should be an irreducible representation of the Poisson algebra
$\F(P)$ of observables of the classical phase space $P$, which should
act in a Hilbert space ${\cal H}$ which is also constructed  
in a natural manner out of the classical system. 
Thus, with any function $f:P\longrightarrow \Re$, it should be associated  
a linear self-adjoint operator $\widehat{f}$,
which acts on ${\cal H}$ and such that, 

\be \widehat{\left\{f,\>g\right\}}= [\widehat{f},\>\widehat{g}]
,\>\forall f,g\in\F(P)\label{rep}\ee
It is well known that this program cannot be fully executed because 
obstructions arise, mainly due to ordering
problems, which prevent the whole $\F(P)$ from being represented. These 
obstructions are not a major problem if one is able a) 
to represent a subset of $\F(P)$ which is big enough to generate the whole
$\F(P)$, and b) to obtain without ambiguities the basic
observables of the theory such as the Hamiltonian ($\equiv$ quantum
temporal evolution), the quantum angular momentum operators, etc.  

Given a classical phase space with Poisson bracket 
$\{,\}$ ($\equiv$ simplectic form $\omega$), with any $f\in\F(P)$  
we associate a natural operator
$X_f:\F(P)\longrightarrow\F(P)$, defined through: 

\be X_f(g)=\{f,\>g\},\>\forall g\in\F(P)\ee
Because of the Jacobi identity, 
these operators also fulfil eq. (\ref{rep}).
These relationships give us a basic guide to the expected nature 
of the Hilbert space of
the quantum theory, ${\cal H}\sim\F(P)$, 
and the quantum operators: $\widehat{f}\sim X_f$. 
The difficulty is that the correspondence 
$f\rightarrow X_f$ is not faithful because the
constant functions are in its kernel. To overcome this problem a
new term has to be added to the 
operators $X$ so as to associate the natural constant
operators with the constant functions. This is achieved by
(non-trivially) enlarging $P$ with a new parameter 
$\zeta\in U(1)$ to give rise to a new
manifold $Q_P$ $-$which is called a {\bf quantum manifold}$-$ 
with a structure of $U(1)$ principal bundle over $P$,  
so that $Q_P/U(1)=P$. 
The dependence of the wave functions with respect to the new co-ordinate 
$\zeta\in U(1)$ is fixed by means of the condition 

\be\Psi(\zeta p)=\zeta\Psi(p), 
\quad\forall \zeta\in U(1)\label{zetafunction}\ee
If $X_\zeta$ is the vector field which 
generates the action of $U(1)$ on $Q_P$, the
constraint (\ref{zetafunction}) reads: 

\be X_\zeta\Psi = i\Psi\label{zetafunction2}\ee
This condition together with the natural requirement 
that the constant functions 
must be properly represented implies that 
the new (pre-)quantum operator associated with 
$f\in\F(P_Q)$ has the (local) expression: 

\be \tilde{X}_f=-i\left[X_f -\left(\hbox{i}_{X_f}\lambda -
if\right)X_\zeta\right]\label{tildeX}\ee
where $\lambda$ is a symplectic potential to $\omega$. 

Let now $\Theta$ be the connection $1$-form 
on $Q_P\longrightarrow P$, which is defined by the conditions  
$\hbox{i}_{X_\zeta}\Theta =1,\>\hbox{i}_{X_\zeta}\d\Theta =0$ and 
$(Q_P,\>\d \Theta)/U(1)\sim (P,\>\omega)$. Then, the operators 
$\widetilde{X}_f$ will be defined by the relationships: 

\be \hbox{i}_{\widetilde{X}_f} \Theta = f, \quad \hbox{i}_{\widetilde{X}_f}
\d \Theta = -\d f\label{tildeX2}\ee 
(This relationships imply in particular that $\hbox{L}_{\widetilde{X}_f}
\Theta =0$.) 

With this procedure, we make sure that the correspondence  
$f\longrightarrow \widetilde{X}_f$ 
is faithful. However, it will in general be reducible: there are 
non-trivial operators, $\widetilde{X_a},\>a\in I$, which commute 
with the basic ones of the
representation, $\widetilde{X}_{q^i},\>\widetilde{X}_{p_j}$. 
The irreducibility has to be achieved by imposing further 
that (some of) these operators act trivially 
on the physical Hilbert space: 

\be \widetilde{X_a}\Psi=0,\quad \hbox{for some}\> a\in I,\>\forall
\Psi\in{\cal H}\label{gqcontraint}\ee  
This last condition roughly amounts to requiring that the wave functions
depend only on the $q^i$'s or the $p_j$'s 
(or a particular combination of these such 
as the creation/annihilation operators). 

\subsection{The GAQ formalism over a connected Lie group} 
The GAQ formalism was originally conceived \cite{[JMP]} 
to improve GQ by freeing it from several limitations and technical
obstructions. Among them we point out the impossibility of considering
quantum systems without classical limit, the lack of a proper (and naturally
defined) Schr\"odinger equation in many simple cases and the ineffectiviness 
in dealing with anomalous systems \cite{[Anomalias]}. 

The main ingredient which enable GAQ to avoid 
these limitations is a Lie group
structure on the manifold $\tilde{G}$ replacing 
the quantum manifold $Q_P$ of
GQ. $\tilde{G}$ is also a principal bundle 
with structure group $U(1)$, but now
$\tilde{G}/U(1)$ is not forced to wear 
a symplectic structure. This way, 
non-symplectic parameters associated with 
symmetries like time translations, 
rotations, gauge transformations, etc. 
are naturally allowed and give rise to 
relevant operators (Hamiltonian, angular momentum, null charges,
etc). Needless 
to say that the requirement of a group structure in $\tilde{G}$ represent some 
drawback, although it is lesser, in practice, 
than it might seem. In particular  constrained quantization (see below
and ref. \cite{[frachall]}) as well as higher-order 
polarizations \cite{[higherpjmp],[marmo]} allow GAQ to be applied to phase 
spaces that do not wear a group structure, thus greatly
expanding the range of applicability of the formalism.

Nonetheless, we  should remark that the GAQ 
formalism is not meant to quantize
a classical system (a phase space) but, rather, 
the quantizing group is the primary quantity and in some 
cases (anomalous groups \cite{[marmo]}, for instance) it is unclear how to
associate a phase space with the quantum theory obtained.

As a general rule, and roughly speaking, 
$\tilde{G}$ is a central extension of a
group $G$ which represent a phase space 
enlarged by the (usually semi-direct) action
of additional (non-symplectic) symmetries. 
As mentioned in the Introduction, GAQ
proceeds associating  quantum systems with already 
given groups $\tilde{G}$, but also
the possibility exists of looking for an appropriate group $\tilde{G}$ out of
a given (classical) Lagrangian ${\cal L}$. 
In this case the solution manifold of
${\cal L}$ (as a phase space) should be the 
starting point to construct the manifold
of $\tilde{G}$.  

The basic structure in the GAQ formalism, 
is, therefore, a Lie group $\widetilde{G}$ 
(see next section where generalizations are discussed) 
which is called the {\bf quantizing group}.   
In this group, there are naturally defined left-invariant 
(right-invariant) vector fields, $\widetilde{X}^L_i$ $(\widetilde{X}^R_i)$ 
as well as left-invariant (right-invariant) forms ${\theta^L}^i$ 
(${\theta^R}^i$). 
As in Geometric Quantization a major role is played by the 
left-invariant form, ${\theta^L}^\zeta$, which is dual of 
the generator of the central subgroup $U(1)$ after a basis 
of the Lie algebra has been chosen. 

\Def: The 1-form $\Theta\equiv{\theta^L}^\zeta$ dual to the vertical 
generator $\widetilde{X}_\zeta$ is called {\bf quantization form}. 

The space of wave functions will now be constructed on the functions on
$\tG$ which fulfil the condition of being $U(1)$-functions, 
which is now written: 

\be \Xi\Psi=i\Psi,\quad \forall \Psi\in \F\left(\tG\right)\ee 
where $\Xi=\widetilde{X}^L_\zeta=i\zeta\frac\partial{\partial\zeta}
=\widetilde{X}^R_\zeta$. 

The quantum operators are the {\it right-invariant}
vector fields. 

Now there are two main points 
to be taken into account: 

a) Some of the parameters of the group are not symplectic; that is, 
there are left invariant vector fields $X^L_i$ such that 

\be \hbox{i}_{X^L_i}\Theta=0=\hbox{i}_{X^L_i}\d \Theta\label{gaq5}\ee

b) The left-invariant and right-invariant vector fields commute. 
Therefore, the right-invariant vector fields do not provide an
irreducible representation of $\tG$ when acting on the space 
of $U(1)$-functions. 

\Def: Let $\widetilde{{\cal G}}$ be the Lie algebra of $\tG$. 
The {\bf characteristic 
subalgebra} ${\cal C}$ of $\widetilde{{\cal G}}$ 
is the subalgebra which is 
expanded by the vector fields which fulfil eq. (\ref{gaq5}). 

\Def: We shall say that a left subspace ${\cal S}$ 
is {\bf horizontal} iff 

\be  \hbox{i}_{X^L}\Theta=0, \qquad \forall   X^L\in  {\cal S}\ee 

\Def: A {\bf polarization subalgebra} ${\cal P}$ 
is a maximal horizontal subalgebra of $\widetilde{{\cal G}}$ such that
${\cal C}\subset{\cal P}$. 

Points a) and b) are taken into account together by imposing 
the polarization conditions on the wave functions: 

\Def: A wave functions $\Psi$ is said to be {\bf polarized} iff  

\be \widetilde{X}\Psi=0,\quad 
\forall\widetilde{X}\in {\cal P}\label{polarized}\ee 
where ${\cal P}$ is a polarization. 
With this requirement, and in the absence of constraints  (see below),  
the quantization procedure is completed if we further 
specify a ${\tilde G}$-invariant integration measure. This 
measure has, in practice, turned out to be derivable from the natural 
one ${\theta^{L}}{}^1\wedge {\theta^{L}}{}^2\wedge...$ on ${\tilde G}$,  
though the general case has not yet been addressed.  
The physical Hilbert space $\H$ is then expanded by the integrable 
polarized wave functions. The physical operators 
are the right-invariant vector fields acting in 
this space and they are unitarily
represented. 
 
\vskip5mm
\noindent{\bf Gauge subalgebra}
\vskip3mm

\Def: We shall say that a right-invariant vector field $\widetilde{X}^R$
is {\bf gauge} if 

\be \hbox{i}_{\widetilde{X}^R}\Theta=0 
\label{gaqgauge}\ee
The subalgebra expanded by all the 
gauge vector fields will be denoted ${\cal N}$ and
will be termed {\bf gauge subalgebra}. 

Since for all $\widetilde{X}^R$ and ${\theta^L}$, 
 $L_{\widetilde{X}^R}{\theta^L}=0$, eq. (\ref{gaqgauge}) implies 
$\hbox{i}_{X^R}\d\Theta=0$. This agrees with the usual description
of the gauge symmetries as the ones which are generated by 
vector fields in the kernel of the presymplectic 
$2$-form (see, for instance, \cite{[reduced]} and references therein). 
Also, in the GAQ formalism, the conserved (Noether) charge associated 
with $\widetilde{X}^R$ corresponds to $\hbox{i}_{\widetilde{X}^R}\Theta$. 
Therefore, the definition above is consistent with the well-known 
fact that gauge symmetries have null conserved
charges (see, for instance, ref. \cite{[Jackiw]} for a direct proof). 

\Prop:  Let ${\cal N}$ be the subspace expanded by the gauge vector 
fields. Then ${\cal N}$ is an ideal of $\widetilde{{\cal G}}$. 

\Proof:  It follows inmediately by making use of the equality
$i_{[X,Y]}= L_Xi_Y-i_YL_X$. 

For $\widetilde{X}^R$ gauge, 
$\widetilde{X}^R\in \hbox{Ker}\Theta\cap\hbox{Ker}\d
\Theta={\cal C}$. Since ${\cal C}$ is expanded by the 
characteristic subalgebra, $\widetilde{X}^R$ must be of the form

\be \widetilde{X}^R 
= \sum_{j\in c} {f}^j \widetilde{X}^L_j\label{gaugeinc}\ee 
Therefore the polarized wave functions are automatically gauge
invariant: 

\be \widetilde{X}^R(\Psi)=0, \quad \forall 
\>\>\widetilde{X}^R\>\hbox{gauge} \ee
and no new (right) constraints need to be imposed.  

\section{The (algebraic) GAQ formalism over a group} 

In this section the GAQ formalism will be presented in a pure 
algebraic language. That is, we shall make use of finite 
quantities and algebraic operations only: 
composition of group elements, subgroups, etc.  
A (desired) consequence of this reformulation 
is that nowhere it is
needed a differential structure on the quantizing group, 
that is, now $\tG$ need not to be a Lie group. 
It can be a discrete or even finite group. 

We shall consider only the case in which the quantizing group 
$\tG$ is provided with a {\it central} 
subgroup $T_0$ which, in this paper,  
will be called {\bf canonical subgroup}. Natural extensions of the 
formalism to more general cases have already been 
discussed in the literature (see for instance \cite{[FormalGroups]}) 
but will not be considered here. 

The canonical subgroup is the centre of
gravity around which the group 
quantization formalism is formulated. 

The GAQ formalism requires us to singularize, appart from the canonical
subgroup, two other subgroups of $\tG$: the characteristic subgroup 
and the polarization subgroup. In addition, the gauge subgroup 
is also naturally defined.   

\Def: We shall say that a subgroup $H\subset\tG$ is {\bf horizontal} 
if $H\cap T_0=\{1_{\tG}\}$, where $1_{\tG}$ 
is the neutral element of $\tG$. 

\Def: Given $g,\>g'\in\tG$, we define the {\bf commutator} of $g,\>g'$
as $[g,\>g']=gg'g^{-1}{g'}^{-1}$. If $S,\>S'$ are two subsets (not
necessarily subgroups) of $\tG$, then $[S,\>S']\equiv\{[g,\>g']\>/g\in
S,\>g'\in S'\}$. 

\Def: The {\bf characteristic subgroup} $C$ of $\tG$ is the   
maximal horizontal subgroup such that 
$[C,\>\tG]\cap T_0=\{1_{\tG}\}$.  

\Def: A {\bf polarization subgroup} $P$ is a maximal horizontal 
subgroup of $\tG$ such that $C\subset P$. 

\Def: The {\bf gauge subgroup} $N$ of $\tG$ is the maximal  
horizontal normal subgroup of $\tG$. 

\Note: Since $N$ is horizontal and $[\tG,\>N]\subset N$,  
then $N\subset C$. 

When $\tG$ is a Lie group the above definitions lead to the ones 
for the Lie algebras in the previous section. 
In particular, because of the following proposition, 
which is the reciprocal of Proposition 2.1, 
the Definition 3.5 corresponds to the one for a gauge subalgebra: 

\Prop: Let $H$ be a horizontal normal subgroup of a Lie 
quantizing group $\tG$ and let $\widetilde{X}_i^R$ be the right 
invariant vector fields which generate $H$, then   

\be \hbox{i}_{\widetilde{X_i}^R}\Theta= 0.\ee 

\Proof: Consider any function  
 $\Psi:\tG\longrightarrow C$ such that 
$\Psi(gh)=\Psi(g)$ for all $g\in\tG,\>\>h\in H$. Then, 
because $H$ is normal,  $\Psi(hg)=\Psi(g)$ also. 
This fact requires that, at any $g\in\tG$, any  
right-invariant vector field $\widetilde{X}_i^R$ 
which generates the left action of $H$ 
can be expressed as a linear combination of 
the left-invariant vector fields $\widetilde{X}_j^L$ 
which only involves the vector fields which generate 
the (right) action of $H$, and the other way round. 
Therefore, since $H$ is horizontal, the charges which are 
associated with the invariant vector fields tangent to $H$  
and to $\Theta\equiv{\theta^L}^\zeta$ are zero. 

\v3mm
The proper quantization proceeds as follows: 

We start with the space $\F(\tG)$ of complex funtions on
$\tG$ and pick up a representation $D_{T_0}$ 
of $T_0$, and 
a right-representation $D_P$ of a polarization $P$,  on $\F(\tG)$.  

\Def: We shall say that $\Psi\in\F(\tG)$ is a 
$D_{T_0}$-function iff   

\be \Psi(zg)=D_{T_0}(z)\Psi(g),
\>\forall g\in\tG,\>\forall z\in {T_0}
\label{DHfunctions}\ee

\Def: A function $\Psi\in\F(\tG)$ is called {\bf polarized} 
($D_P$-polarized) iff  

\be \Psi(gp)=D_P(p)\Psi(g),\>\forall g\in\tG,\>\forall p\in P 
\label{DPpolarized}\ee

In absence of constraints,  these conditions fully 
determines the Hilbert space of
the theory: it is given by the set of all (square integrable) polarized 
$D_{T_0}$-functions in $\F(\tG)$. 
The dynamical operators are all the
elements in $\tG$, and they act as finite 
{\it left} traslations on the Hilbert space: 

\be (\widehat{g}\Psi)(g')=\Psi(g^{-1}g'), 
\forall g,\>g'\in\tG\label{finiteops}\ee 

Therefore the gauge subgroup, which corresponds to gauge constraints which 
have been solved classically, is automatically and trivially represented. 

\v5mm 
\noindent{\bf Constraint quantization and good operators}
\v3mm 

As is well known (see basis references in \cite{[Teitelboim]}, 
see also \cite{[Henneaux]}),  
there is a close relationship between constraints and gauge
symmetries. Loosely speaking, the existence of a gauge symmetry 
suffices to have a constrained system, and first-class 
constraints generate gauge symmetries. 
Constraints are not, however, always due to the presence of 
gauge symmetries in the system: 
the former are more general than the latter. 

Here we shall consider only   
the case in which the constraints 
close into a subgroup $\widetilde{T}\subset\tG$.   
The constraint subgroup $\widetilde{T}$ is required to be a 
fibre group of $\tG$, i.e., $\tG\longrightarrow \tG/\widetilde{T}$ 
is a principal bundle and to contain $T_0$ as a fibre group, 
i.e. $\widetilde{T}\longrightarrow \widetilde{T}/T_0$ 
is also a principal bundle. 
In particular, $\widetilde{T}$ should be regarded as a quantizing group 
with the same canonical subgroup $T_0$ as $\tG$. 

When there are constraints the procedure described above  
has to be completed with additional conditions on the wave
functions. Now the physical Hilbert space is made up of all the 
polarized $T_0$-functions which are constrained: 

\Def. A wave function $\Psi:\widetilde{G}\longrightarrow C$ 
is termed {\bf constrained} iff 

\be \Psi(t*g)= D_{\widetilde{T}}(t)\Psi(g),\quad 
\forall t \in \widetilde{T},\> g\in\widetilde{G}\label{constrainedPsi}\ee 
where $D_{\widetilde{T}}$ is an irreducible representation of 
$\widetilde{T}$. 

Representations of $\widetilde{T}$ which are compatible with $D_{T_0}$ 
are naturally found by applying 
the GAQ formalism to $\widetilde{T}$.  We, therefore, need  
the same collection of subgroups of $\widetilde{T}$ in relation to $T_0$ 
as just described for $\widetilde{G}$. 
When there is danger of confusion, these subgroups of $\widetilde{T}$ 
will be signalled by placing a prefix $\widetilde{T}$ before them. 
Thus, we shall have the $\widetilde{T}$-characteristic subgroup, 
the $\widetilde{T}$-polarization subgroup and so on. 

Clearly not all the operators in $\tG$ will  
preserve the representation $D_{\widetilde{T}}$ of $\widetilde{T}$; 
for the dynamical operators that do we shall use the 
name {\bf good operators} \cite{[CMP]}. 
The group of all the good operators thus constitutes 
the natural generalization 
of the concept of normalizer of $\widetilde{T}$.  
This is the manner in which the concept of gauge subgroup (gauge 
symmetries) is incorporated into the quantum level. 

In some cases (where $\widetilde{T}$ is connected and is  
not a direct product $\widetilde{T}\neq T_0\otimes T$) the 
$\widetilde{T}$-function condition (\ref{constrainedPsi})  
may not be compatible with the representation $D_{T_0}$ for $T_0$. 
Then we must soften that requirement and consider, 
rather than the whole $\widetilde{T}$, 
a subgroup $T_0\oplus P_T$, 
where $P_T$ is a polarization subgroup of 
$\widetilde{T}$. This subtlety does not arise, however, in the 
models we shall consider in the present paper, in which the 
whole $\widetilde{T}$ can be represented in a way compatible 
with the $T_0$-function condition. 

When $\widetilde{T}$ is a non-trivial 
central extension, it is sometimes said that 
the gauge symmetries are ``anomalous''. Nonetheless, these ``anomalies'' 
do not necessarily imply obstruction to 
quantization, and do not particularly when the condition 
(\ref{constrainedPsi}) can be imposed for the entire 
$\widetilde{T}$. 

\newpage 

\v5mm
\bc
\large {\bf PART 2. LINEAR FIELDS}
\ec\normalsize 
\section{Linear fields} 
Throughout this section, we shall consider a theory with fields 
$\varphi^a,\>\> a=1,...,N,$ and action 

\be S=\int_{\cal M}\mu\,
\L(\varphi^a,\partial_\mu \varphi^a)\label{1.1}\ee 
The space-time manifold ${\cal M}$, 
with volume element $\mu=\d^{D+1}x$, will
always be homeomorphic to $\Sigma\times\Re$ where $\Re$ represents the
time-like directions and $\Sigma$ is any ($D$-dimensional)
spacelike hypersurface.  When picking up a particular Lagrangian, we 
shall make use, if necessary, of the 
indetermination under a total divergence. 

The set of all fields, irrespective of whether or not they satisfy
the Euler-Lagrange equations of motion will be denoted by $\F$. 
We shall term any solution of the (classical)
equations of motion as {\bf trajectory}, or classical trajectory.  
${\T}$ will be the set of all the trajectories of the
system. 

If a (classical) theory of fields, 
$(S,\>\F)$ is linear, the
space ${\T}$ of all the solucions of the equations of motion is a vector
space. That is, if $\varphi$ and $\phi$ are solutions, 
so is $\lambda\varphi+\beta\phi$ for any $\lambda,\beta\in \Re$. 
Therefore ${\T}$ can be regarded as an (abelian) group 
of symmetries of the theory with composition law:

\be \varphi'' = \varphi' + \varphi\label{basicgroup}\ee
This group will be denoted as $G_{(\varphi)}$. 

\Theorem: If $(S,\>\F)$ is a classical 
theory of linear fields,
with Euler-Lagrange equations of motion 
$\left([E-L]\varphi\right)_a=0$, then 

\begin{tabular}{ll}
{}&{}\\
a)&$\L(\varphi+\phi)= \L(\varphi)+\L(\phi)+ 
\left([E-L]\phi\right)_a \varphi^a 
+ \p_\mu J^\mu(\varphi,\phi),\> \forall \> \varphi,\>\phi\>\in \F$\\ 
{}&{}\\
b)& A Lagrangian is given by:
\end{tabular}

\be \L(\varphi) = \frac12\left([E-L]\varphi\right)_a \varphi^a\nonumber\ee

Therefore, there exists a Lagrangian which vanishes ``on-shell'', i.e. 
$\L(\varphi)=0$ for any classical trajectory $\varphi$. 

\Proof: The point a) follows inmediately 
if we look at $\L(\varphi+\phi)$ as a variation of the
Lagrangian, a variation similar to the one 
which gives the Euler-Lagrange equations of motion. 

If in the equality a) we make $\varphi=\phi=\frac12\kappa$, we obtain:

\be \L(\kappa)=\frac12\left([E-L]\kappa\right)_a\,\kappa^a + 
\frac12\p_\mu J^\mu(\kappa,\kappa)\label{LC}\ee 
The new Lagrangian 
\be \widehat\L(\kappa) = \L(\kappa) 
-\frac12\p_\mu J^\mu(\kappa,\kappa)\label{newlag} \ee
fulfils part b). 

\Corollary: Since the current $J^\mu(\varphi,\phi)$ is bilinear, it 
can be chosen to be: 

a) Divergenceless on trajectories. 

\noindent and 

b) Antisymmetric.

\Proof: 

a) This is a consequence of part a) of Theorem 4.1   
if a Lagrangian that vanishes on shell is chossen. 

b) It is sufficient to show that $J^\mu(\kappa,\kappa)$ is identically null. 
If $J^\mu(\kappa,\kappa)$ were not identically null, 
then $\widetilde{J}^\mu(\kappa,\kappa)$, where  
$\widetilde{J}^\mu(\varphi,\phi)=J^\mu(\varphi,\phi) 
-\frac12 J^\mu(\varphi,\varphi) - 
\frac12 J^\mu (\phi,\phi)$ is also an admisible current, would.  
However, both $J^\mu$ and $\widetilde {J}^\mu$ have to be bilinear. 
Therefore, $J^\mu(\kappa,\kappa)=0\>\> \forall 
\kappa$ and $\widetilde{J}^\mu=J^\mu$. 

\Def: The current $J^\mu$ for which a) and b) hold will be called the {\bf 
canonical current} of $(S,\>\F)$. 

\Note: There is, in fact, a shorter but equivalent 
way of obtaining the canonical $J^\mu$. 
If in Theorem 4.1.a) we exchange $\varphi$ and $\phi$ and then antisymmetrize, 
we get: 

\be \p_\mu\left(\frac12J^\mu(\varphi,\phi) - \frac12J^\mu(\phi,\varphi)\right) 
= \frac12\left([E-L]\varphi\right)_a \phi^a- 
\frac12\left([E-L]\phi\right)_a \varphi^a\ee 
The current $\left(\frac12J^\mu(\varphi,\phi) - 
\frac12J^\mu(\phi,\varphi)\right)$ is
the canonical current of $(S,\>\F)$.  

\Corollary: If $\widetilde{\L}=\L+\p_\mu\Lambda$, 
then $\widetilde{J^\mu} =J^\mu$

\Def: For all $\varphi,\, \phi\in\T$ we define the {\bf canonical product} 
$\Omega\left(\varphi,\phi\right)$ by means of:

\be \Omega\left(\varphi,\phi\right) = \int_\Sigma\d \sigma_\mu\>
J^\mu(\varphi,\phi)\label{sproduct}\ee
where $\Sigma$ is any Cauchy hypersurface in ${\cal M}$. 
Therefore it is bilinear, antisymmetric and 
independent on the $\Sigma$ hypersurface. 

The canonical product of two solutions  
$\varphi$ and $\phi$ is nothing other than  
the Noether charge associated with the 
symmetry generated by $\varphi$ 
in the point $\phi \in {\cal T}$, or (minus) the other way round. 
It measures the degree to which 
the classical trajectories $\varphi,\phi$ 
are coordinate-momentum conjugate to each other. 

\Note: Notice that the potential current of the theory is $j^\mu = 
J^\mu(\varphi,\delta \varphi)$. The symplectic form is therefore given by 
\cite{[cps],[reduced]}:  

\be \omega = -\int_{\Sigma}\d \sigma_\mu \delta j^\mu\label{w1}\ee

\Theorem: With $\Omega$ defined above, the following composition
law defines a central extension of $G_{(\varphi)}$ 
which will be denoted $\tG_{(\varphi)}$: 

\bea \varphi''(x) &=& \varphi'(x) + \varphi(x)\label{gbasic1}\\
\zeta''&=&\zeta'\zeta\exp{i}\Omega\left(\varphi',\varphi\right) 
\label{ext1}\eea
where the fields $\varphi,\>\varphi'...$ are trajectories and 
$\zeta, \zeta'...\in U(1)$.  

\subsection{Space-time and internal symmetries} 
In addition to the symmetries in $G_{(\varphi)}$, which act
additively, there are in general other symmetries, such as space-time
or internal ones, which act multiplicatively. 
In this section, we shall
study the conditions under which the group $\tG_{(\varphi)}$ can be
enlarged with these other symmetries. 

First of all we note that since the composition 
of two symmetries is another symmetry, any two groups of symmetries 
$U_1$ and $U_2$, can be enlarged to obtain a new group
$U_3$ such that $U_1,\,U_2\subset U_3$.  
Therefore, without loss of generality, we can 
consider a single group of symmetries
$U=\left\{u,v,..\right\}$. The requirement of being symmetries
is that, if $\varphi\in\T$, then $u(\varphi)\in\T$. 

These symmetries (which should be thought of as 
being like $SU(2)$, the Poincar\'e, the conformal or 
the Virasoro groups) usually act on $\F$ ($\T$) through a previous 
representation in the space-time. 

For any field $X$ which generates the action of $U$ on $\F$, we have 

\be L_X\L\mu = \d \Lambda_X\label{lagsym}\ee 
with $\Lambda_X$ a space-time $D$-form. 

Eq. (\ref{lagsym}) together with Corollary 4.1.2 imply that   
the following lemma holds:  

\Lemma: Let $U_0$ be the component of $U$ which 
is connected to the identity, then
$\Omega\left(u(\varphi),u(\phi)\right)= 
\Omega\left(\varphi,\phi\right),\,\forall
\varphi,\phi\in\T,\> \forall u\in U_0$.  

For symmetries which are not connected to the identity, such as parity or 
temporal inversion, this lemma has to be relaxed, as we can have 
{\bf anticanonical symmetries}, that is, symmetries $u$ for which 
$\Omega\left(u(\varphi),u(\phi)\right)=-\Omega\left(\varphi,\phi\right)$. In
general, the action of $U$ on $\Omega$ defines a representation
$\epsilon$ of $U$ on $Z_2=\{+,-\}$. Then we shall have:   

\Theorem: With the fields as defined above, the following composition
law is a group:

\bea u''&=&u'*u,\quad u,\>u',\>u''\in U\nonumber\\
\varphi''(x) &=& \varphi'(x) + \left(u'(\varphi)\right)(x),
\quad \varphi,\>\varphi',\>\varphi''\in \T\label{gtotal}\\
\zeta''&=&\zeta'\zeta^{\epsilon(u')}\exp{i}
\Omega\left(\varphi',\left(u'(\varphi)\right)\right),\quad  
\zeta,\>\zeta',\>\zeta''\in U(1)\nonumber\eea
This group will be denoted $\tG_{(S,\>\F)}$. Note that when there are 
anticanonical symmetries in $U$, it is no longer a central extension.  

For the sake of brevity we shall consider only canonical symmetries; 
that is, symmetries for which $\epsilon(u)=1$. 
Anticanonical transformations, which give rise to  
interesting subtleties, will be the subject 
of a separate study \cite{[anticanonical]}. 

Depending on the context, 
expressions for the group $\tG_{(S,\>\F)}$ which are different from  
eq. (\ref{gtotal}) --where the 
symmetry group $U$ acts from the left--  
and which are obtained from it  
by means of a change of variables, may appear to be  
more natural ones. For instance 

\bea u''&=&u'*u,\quad u,\>u',\>u''\in U\nonumber\\
\varphi''(x) &=& \left(u^{-1}(\varphi')\right)(x) + \varphi(x),
\quad \varphi,\>\varphi',\>\varphi''\in \T\label{gtotal2}\\
\zeta''&=&\zeta'^{\epsilon(u)}\zeta\exp{i}
\Omega\left(\left(u^{-1}(\varphi'),\>\varphi\right)\right),\quad  
\zeta,\>\zeta',\>\zeta''\in U(1)\nonumber\eea
where the symmetry group $U$ acts on the left instead. 
In the rest of this paper, we shall make use of combinations of these two
presentations in which some subgroups of $U$ act from the left and others 
from the right. 

\v3mm
\noindent{\bf Example: the non-relativistic free particle and the
Galilei group}
\v2mm  

\noindent As a first example of the construction above, let us 
consider the non-relativistic free particle --regarded as a 
$(0+1)$-dimensional field theory -- and  
construct the quantizing group for it. 
In spite of its simplicity, we follow the
same steps as for a standard field in contrast with the  
quantum-mechanical treatment of the 
free particle \cite{[JMP]}. For more examples, 
see below and ref. \cite{[config]}
where the harmonic oscillator, which provides an useful link
between mechanics and field theory, is also considered. 

A Lagrangian for the non-relativistic free-particle is:

\be L'_{FP}({\bf x})= \frac{m}2\dot{\bf x}^2\label{fp}\ee
We have 
\be L'_{FP}({\bf A}+{\bf B})= \frac{m}2\left[\dot {\bf A}^2 + 
\dot {\bf B}^2 - 
2\ddot {\bf B}{\bf A} +
2\frac\d{\d t}\left(\dot{\bf B}{\bf A}\right)\right]\ee
Thus, the associated on-shell-vanishing Lagrangian, 
the equations of motion and 
the canonical product are, respectively:   

\bea L_{FP}&=&-\frac{m}2\ddot{\bf x}{\bf x}\nonumber\\
-\frac{m}2\ddot{\bf x} &=& 0\nonumber\\ 
\Omega_{FP}\left({\bf A},\>{\bf B}\right)
&=&\frac{m}2\left[{\bf \dot AB-A\dot
B}\right]\label{OmegaFP}\eea 

Now we can consider the spatial rotations and time translations as the 
group of space-time symmetries. 
These act on $\F_{FP}$ as follows

\bea (R{\bf A})^i(t)&=&{R^i}_jA^j(t),\> {R^i}_j\in O(3)\nonumber\\
(T_b({\bf A})(t)&=&{\bf A}(t-b), \quad b\in \Re\label{FP2}\eea 

Now the general solution to the equations of motion is: 

\be {\bf x}(t)={\bf Q} + {\bf V}t,\quad {\bf Q},
\>{\bf V}\in \Re^3\label{solFP}\ee
and ${\bf Q},\>{\bf V}$ can be taken as the coordinates in $\T_{FP}$.  
It is simple to see that 

\bea R({\bf Q})^i&=&{R^i}_jQ^j,\qquad R({\bf V})^i={R^i}_jV^j\nonumber\\
T_b({\bf Q})&=& {\bf Q}-{\bf V}b,\qquad T_b{\bf V}={\bf V}\nonumber\eea 
The group $\tG_{FP}$ is therefore given by

\bea b''&=&b'+b\nonumber\\
{\bf Q}''&=&{\bf Q'} +{\bf V'}b +R'\left({\bf Q}\right)\nonumber\\
{\bf V}''&=&{\bf V'} +R'\left({\bf V}\right)\label{galileo}\\
\zeta''&=&\zeta'\zeta\exp\frac{i}2m\left[\left({\bf Q'} +
{\bf V'}b\right)R'\left({\bf V}\right) -  
{\bf V'}R'\left({\bf Q}\right)\right]\nonumber\eea 
which is the Galileo group (extended by 
the Bargmann cocycle \cite{[JMP]}). 

\subsection{Quantization}
Now that we have found out the quantizing group $\tG_{(S,\>\F)}$, we shall   
apply to it the GAQ formalism presented in Part 1. 

To identify the characteristic subgroup, we have to construct
the commutator of two generic elements 
$g=(u,\varphi,\zeta)\in\tG_{(S,\>\F)}\>,\>\>  
g'=(u',\varphi',1)\in C$. $C$ will be the maximal subgroup such that   
$\left[g\>,\>g'\right]=(1_U,0,\zeta)$ implies  $\zeta=1$. 

We have 

\bea
g'g&=&(u'u,u^{-1}(\varphi')+\varphi,
\zeta'\zeta\exp\frac{i}2\Omega(u^{-1}(\varphi'),\varphi))\nonumber\\
g^{-1}&=& (u^{-1},-u(\varphi), \zeta^{-1})\nonumber\\
gg'&=&(uu',u'^{-1}(\varphi)+\varphi',
\zeta\zeta'\exp\frac{i}2\Omega(u'^{-1}(\varphi),\varphi')
\label{commutatortG}\\
g'g(gg')^{-1}&=&\left(u'u(uu')^{-1},uu'[u^{-1}(\varphi')+\varphi]-
uu'[u'^{-1}(\varphi)+\varphi'],\right.\nonumber\\
&&\exp\frac{i}2\left[\Omega(u^{-1}(\varphi'),\varphi)-
\Omega(u'^{-1}(\varphi),\varphi')\right.
\nonumber\\
&&\qquad \left.\left.- \Omega(uu'[u^{-1}(\varphi')+\varphi],
uu'[u'^{-1}(\varphi)+\varphi'])\right]\right)\nonumber\eea 

Therefore, $g'=(u',\varphi',1)$ has to fulfil  
\be \Omega(\varphi',u(\varphi) + u'^{-1}(\varphi))=0\quad 
\forall g=(u,\varphi,\zeta)\in \tG_{(S,\>F)}\ee 
This implies 

\be C=U\oplus N\label{C}\ee   
with $N=\hbox{gauge subgroup}=\{(1_U,\varphi',1)/\Omega(\varphi',\varphi)=0 
\quad \forall g=(u,\varphi,\zeta)\in \tG(u,\varphi,\zeta)\}$. 
[$U\oplus N$ stands for the subgroup generated 
by $U\cup N$ and it also means $U\cap N=\{1_{\tG}\}$.]

We recall now that a polarization subgroup is a maximal horizontal  
subgroup $P$ such that $C\subset P$. 
Thus, any $P$ is generated by  

\be P=C\cup P_\varphi\label{P}\ee
where $P_\varphi$ is the maximal horizontal 
subgroup  such that $\Omega(v(\varphi),\varphi')=0\quad  \forall 
g=(1_U,\varphi,1)\>,\hfill\break 
\>g'=(1_U,\varphi',1)\}\in P_\varphi,\quad\forall v\in U$

\Def: A {\bf Lagrangian subgroup} is any 
subgroup $L=\{(1_U,\varphi,1)\}$ such that 
$\Omega\left(\varphi,\>\varphi'\right)=0$, for any $(1_U,\varphi,1),\> 
(1_U,\varphi',1)\in L$. 
If $U(L)\subset L$ it will be called 
{\bf invariant Lagrangian subgroup}. 

We, therefore, have:

\Prop: Any polarization subgroup $P$ is generated by 
$U\cup N\cup L$, where  
$L$ is a maximal invariant Lagrangian subgroup.  

\subsection{Holomorphic quantization}

We now consider the case when there are two subgroups $L,\>
\bar{L}\subset\tG$ which fulfil  

a) $\bar{L}$ is a Lagrangian subgroup (not necessarily invariant),  

b)  ${L}$ is an invariant Lagrangian subgroup, 

c)  $\tG_{(S,\>\F)} = U\oplus L \oplus\bar{L}\oplus U(1)$. 

Therefore, any trajectory $\varphi$ has a unique decomposition 

\be \varphi= a +\bar{a}, \quad \hbox{where}\quad (1_U, a, 1)\in L, \> 
(1_U, \bar{a}, 1)\in \bar{L}\label{PmasbarP}\ee

\Note: In general, to find $L$ and $\bar{L}$ with
the properties above, it is necessary 
to go to $\bar{\F}$, the complexified $\F$,    
and to consider instead the group 
$\tG_{(S,\>\bar{\F})}\supset\tG_{(S,\>\F)}$ 
over that complexified space. In this case, the third condition above
takes the form: 

c') $\tG_{(S,\>\F)}\subset U\oplus L \oplus\bar{L}\oplus U(1)  
= \tG_{(S,\>\bar{\F})}$. 

If we take $\varphi=\bar{a} + a$, the polarization
$P=U\oplus L$, and we pick up the trivial representation for it, 
one of the $D_P$-polarization conditions reads: 

\be \Psi(u,\bar{a}+a, \zeta \exp i\Omega(\bar{a},\>a))=
\Psi(u,\bar{a},\zeta)\ee 
This equality,  together with the $U(1)$-function 
condition on $\Psi(u,\varphi,\zeta)$, implies: 

\be \Psi(u, \varphi, \zeta)=
\zeta\Phi(u,\bar{a})\exp[-i\Omega(\bar{a},\>a)] \ee 
The rest of the polarization conditions reads: 

\be \Psi(u'u,u^{-1}(\varphi), \zeta)=
\Psi(u,\varphi,\zeta)\ee 
Therefore 
\be \Phi(u'u,u^{-1}(\bar{a})) = \Phi(u',\bar{a})\label{finiteSchr}\ee 
where we have made use of the fact that $L$ is an invariant
Lagrangian subgroup. Since $\bar{L}$ may not be invariant, 
$u^{-1}(\bar{a})$ is not in general in $\bar{L}$ 
 However, whatever the case,   
eq. (\ref{finiteSchr}) gives the (finite)  
action of the space-time and internal symmetries in the wave functions.
The infinitesimal action, and in particular the Schr\"odinger
equation, can be obtained as the first-order terms 
in the power series in the parameters of the symmetries. 

In the quantum theory of relativistic fields a splitting which
fulfils the requirements above -- and where both $\bar{L}$ and $L$ are
invariant under the (proper) Poincar\'e group --  
is the usual one into negative- and 
possitive-frequency parts. On the other hand, 
the non-relativistic free particle provides an   
interesting and simple example in which the trajectories 
${\bf x}$ split as ${\bf x}={\bf a} +\bar{{\bf a}}$ where 
${{\bf a}}$ is invariant under $U$  whereas $\bar{{\bf a}}$ is not. 
Here $U$ is generated by the time translations and the spatial rotations, 
the trajectory ${{\bf a}}$ is defined by
${{\bf a}}(t)= {\bf x}(t_0)$ and the trajectory $\bar{{\bf a}}$ 
is defined by $\bar{{\bf a}}(t)={\bf x}(t)- 
{\bf x}(t_0)$,   
for all $t\in \Re$ and a fixed $t_0\in \Re$.  This splitting  
corresponds to the familiar parametrization of the phase space with  
position and momenta. The fact that the subspace of positions --
that is, the subset of trajectories with null momentum-- is
invariant whereas the one of momenta -- that is, the 
subset of trajectories with null initial position -- 
is not invariant only apparently contradicts 
the usual transformation of the corresponding 
classical and quantum operators. 

\section{The Maxwell theory in Minkowsky space}
\label{Maxwell} 

From here on in the present paper 
we shall ilustrate over the Maxwell field and the 
abelian Chern-Simon models some aspects of the GAQ formalism 
we have theorized about in the previous sections. The quantization of
the electromagnetic field has been carried further in several papers. In
particular, refs. \cite{[BRST]} and \cite{[empro]} can both 
be regarded as natural continuation of the present
section. Ref. \cite{[fieldsJPA]}, where the Klein-Gordon field as well
as the Proca field are quantized, may also be consulted. 

The usual action for the Maxwell field is: 

\bea S'_{em}=\int\>\d^4x\>\left\{ 
-\frac14 F_{\mu\nu}\>F^{\mu\nu}\right\}\label{em0}\eea 
where 

\be F_{\mu\nu}=\p_\mu A_\nu -\p_\nu A_\mu \label{em1}\ee

It is, however, more natural, and the best for our purposes, to
consider $F^{\mu\nu}$ and $A_\mu$ as independent fields, related only 
by the (now {\it equations of motion}) eq. (\ref{em1}). 
The action which mirrors this point of view is: 

\be S_{em}= \int\>\d^4x\>\left\{ \frac14 F_{\mu\nu}\>F^{\mu\nu}
-\frac12F^{\mu\nu}\left(\p_\mu A_\nu - 
\p_\nu A_\mu \right)\right\}\label{em2}\ee 

As is well known, the Maxwell action is invariant under the 
conformal group, which is made up of compositions of 
the following operations on the space-time:
        
$\ba{ll} \hbox{a) Space-time translations:}& (ux)^\alpha = x^\alpha
+ a^\alpha\\    
\hbox{b) Lorentz transformations:}&(ux)^\alpha=\Lambda^\alpha_\mu
x^\mu\\
\hbox{c) Dilatations:}& (ux)^\alpha=\e^\lambda x^\alpha\\
\hbox{d) Special conformal transformations:}&  
         (ux)^\alpha=\frac{x^\alpha +c^\alpha x^2}
        {1+2cx+c^2x^2}\ea$
\v3mm

The quantizing group for the electromagnetic group is therefore 
\cite{[config]} 

        \bea\label{m1}
        u''&=&u'*u\qquad\hbox{Conformal (sub)group}\nonumber\\
A''_\mu(x)&=&
        \frac{\p u^{\alpha}}{\p x^\mu}A'_\alpha(ux)+A_\mu(x)\\
        &\equiv&(S(u^{-1})A')_\mu(x)+A_\mu(x)\nonumber\\
F''_{\mu\nu}(x)&=&\frac{\p u^{\alpha}}{\p x^\mu}
\frac{\p u^{\beta}}{\p x^\nu}F'_{\alpha\beta}(ux)+F_{\mu\nu}(x)+\\
        &\equiv&(S(u^{-1})F')_{\mu\nu}(x)+F_{\mu\nu}(x)\nonumber\\        
\zeta''&=&\zeta'\zeta\exp i\Omega_{em}
\left(S(u^{-1})(A'),\>A\right)\eea
where $S$ is the representation of the conformal group that acts 
on the electromagnetic  vector field. This action is the 
natural one and means that the potential vector has null 
conformal weight.  

The canonical current is 
\bea\label{Jem}
        {\cal J}_{em}^\mu\left(g',g\right)(x)
     &=&\frac12\left[F'^{\mu\nu}(x) A_\nu(x)-
        A'_\nu(x) F^{\mu\nu}(x)\right]\eea 

\subsection{Non-covariant approach}

Let us write down the action (\ref{em2}) in terms of the 
electric field ${\bf E}$ and the potentials $A_\mu=(A_0,\>{\bf
A})$. In doing so we solve the constraint ${\bf B}={\bf \nabla\times A}$ and
place it back into the Lagrangian. This takes the form 
(save for total derivatives)   

\be \L_M= E^i\dot{A}_i -\frac12\{{\bf E}^2 +({\bf \nabla\times A})^2\}+
A_0\p_i E^i\label{MaxwellEA}\ee
The Lagrangian is constrained with $A_0$ as a Lagrange multiplier 
and constraint 
 
\be \p_i E^i=0 \label{nablaE}\ee
The gauge symmetry of this constrained Lagrangian is the usual one: 
$A_\mu\longrightarrow A_\mu +\p_\mu \Lambda$.  

If space-time symmetries are not considered, 
the quantization of this system with our formalism   
is straightforward -- it amounts to the quantization of three
Klein-Gordon fields in a fixed reference frame -- 
and reproduces the quantum theory of the 
electromagnetic field in the (non explicitly covariant) 
radiation gauge. The quantizing group is   

\bea 
(A''_0 &=& A'_0 + A_0)\nonumber\\
{\bf A}''&=&{\bf A}'+{\bf A}\nonumber\\
{\bf E''}&=&{\bf E}' + {\bf E}\label{bfE}\\
\zeta''&=&\zeta\zeta'\exp\frac{i}2
\int\>\d^3x\sum_{i=1,2,3}\left\{A'_iE^i-{E'}^iA_i\right\}\nonumber\eea
and the subgroup of constraints is 
$\widetilde{T}=\{({\bf A},0,\zeta)/{\bf A}={\bf \nabla}
\Lambda\>\hbox{for some}\>\Lambda\}$. 

\subsection{Covariant gauge fixing, ghost term and bosonic BRST symmetry}

In this section, we construct the quantizing group for the
covariant gauge-fixed Maxwell Lagrangian and 
show how the (bosonic) BRST transformation 
arises as a one-parameter group of internal symmetries  
(in ref. \cite{[BRST]} the present development was carried further; 
see \cite{[empro]} for a thorough an 
unified treatment of the electromagnetic and Proca fields). 

Let us therefore consider the Lagrangian 

\be \L = -\frac14F^{\mu\nu}F_{\mu\nu} - \varphi \p_\mu A^\mu
+\frac{1}{2\lambda} \varphi^2 +\p^\mu c \p_\mu c\label{L2}\ee 
where $\varphi$ is a gauge-fixing Lagrange multiplier and $c$ are
ghost fields. It is straightforward to show that 
this Lagrangian is invariant under 
the following (bosonic BRST) symmetry with parameter 
$\Lambda$: 

\bea \delta A_\mu&=& \Lambda \p_\mu c \nonumber\\
\delta c &=& -\frac12\varphi \Lambda\label{BRSTinf}\\
\delta \varphi&=&0\nonumber\eea 
The finite transformations are given by 

\bea u_\Lambda(A)_\mu &=& A_\mu +\p_\mu c \Lambda -
\frac12\p_\mu \varphi\Lambda^2\\
u_\Lambda(c)&=& c -\frac12\varphi\Lambda\eea 

The general theory shows us that the quantizing group, which includes
the BRST bosonic symmetry but no space-time or internal symmetries, is 
($b\equiv \Lambda$): 

\bea A_\mu'' &=& A_\mu' + A_\mu 
- \p_\mu c'b - \frac12 \p_\mu\varphi' b^2 \nonumber\\
\varphi''&=& \varphi' +\varphi\nonumber\\
c''&=&c'+c +\frac12\varphi'b\label{BRSTgroup}\\
b''&=&b'+b\nonumber\\
\zeta''&=&\zeta'\zeta\exp i \int_\Sigma\d \sigma_\mu J^\mu\nonumber
\eea 
with 

\bea J^\mu &=&\frac12\left(({A'}_\nu - \p_\nu c'b -  
\frac12 \p_\nu\varphi' b^2) F^{\mu\nu}(x)-
A_\nu(x){F'}^{\mu\nu}(x)\right)\nonumber\\
&&+\frac12\left(A^\nu\varphi' -(A'^\nu - \p^\nu c'b - \frac12
\p^\nu\varphi' b^2)\varphi\right)\nonumber\\
&&+\left((c'+\frac12\varphi'b)\p^\mu c -(\p^\mu c'+ 
\frac12\p^\mu\varphi'b)c\right)\label{J2}\eea

Now, if we Fourier transform the fields and make use of the
equation of motion 

\be \varphi=\lambda\p_\mu A^\mu\label{vphipmuamu}\ee  
we shall obtain the group law in ref. \cite{[BRST]}. 

\section{The abelian Chern-Simons theory} 
\label{Chern-Simons}
Let $\M$ be a three-dimensional 
manifold which can be decomposed into the
form $\M=\Sigma\times\Re$ with $\Sigma$ 
an orientable two-dimensional surface. 

The action for an abelian Chern-Simon model is given by \cite{[Witten]}: 

\bea S_{ACS}&=& \frac{k}{4\pi}\int_\M \left(A\wedge \d A\right)\label{cs1}\eea
where $A$ is a one-form with takes values on the Lie algebra ${\cal G}$   
of some abelian lie group $G$ [There is in fact a direct
generalization of the abelian Chern-Simons theories to higher (odd) 
dimensions. In these generalization, $\Sigma$ is a $2D$ manifold  
and $A$ a $D$-form for arbitrary natural number $D$. Many of the
results we present here can be extended to these theories, with 
one-dimensional quantities replaced with higher dimensional ones]. 
It is simple to show that $S_{ACS}$ is
invariant under gauge transformation $A\rightarrow A+\d \Lambda$ for 
any $\Lambda:\M\longrightarrow g$. 

It is straightforward to show that 
the equations of motion and the canonical product are, respectively:   

\bea \d A \equiv F &=&0 \label{cs4}\\
\Omega_{ACS}\left(A',A\right)&=&\int_\Sigma\>J 
= \frac{k}{4\pi}\int_\Sigma\>A'\wedge A \label{cs5}\eea 
Thus, $\T_{ACS}\equiv\F_{{\cal C}}$ where $\F_{{\cal C}}$ 
is the set of all flat connections over $\M$. 

The exterior derivative commutes with the pullback operator 
$\hbox{}^*$. Therefore, if $f$ is a diffeomorphism of $M$ and $A$ and
$A'$ are solutions of the equation of motion (\ref{cs4}),  
then $A'+f^*A$ is also a solution.  

All this, together with the general theory, implies that 
the following composition law defines a group, $\widetilde{G}_{CS}$, 
the quantizing group for the abelian Chern-Simons model:  

\bea f''&=&f'\circ f\>,\quad f,f',f''\in \Diff_0({{\cal M}})\nonumber\\
A''&=&{f^{-1}}^*A'+A\label{tGACS}\\
\zeta''&=&\zeta\zeta'\exp\Omega_{CS}\left({f^{-1}}^*A',A\right)
\nonumber\eea 

The general theory shows that the characteristic subgroup is 
$C_{CS}\equiv N_{CS}= \{(f,A,1)/\break 
A=\d \Lambda\quad\hbox{for some}\quad \Lambda\}$.  
The quantum conditions (\ref{DPpolarized}) imply then that the quantum wave
functions should be functions of topological and gauge invariant
quantities only.  To best deal with these conditions let us  
remind the reader that all the gauge invariant information of a
connection can be extracted from the Wilson loops. 
These are quantities defined by 

\be W(A,\gamma)=\exp\int_\gamma A\equiv A(\gamma)\label{wilsonloops}\ee
for any loop $\gamma$ on ${\cal M}$. 
Therefore (the gauge invariant part of) a connection can be seen as an
application 

\be A:\L_\M \rightarrow G\>/ A(\gamma'\circ\gamma)=A(\gamma')A(\gamma)
\label{compcon}\ee 
where $\L_\M$ is the group of loops on $\M$ [With a slight abuse of 
notation, we shall use the same letter for the connection 1-forms as for the  
applications they define]. Eq. (\ref{cs4}) also implies that the 
diffeomorphisms of ${\cal M}$ which are connected with the identity 
act trivially on the applications $A$. This is not the case with the 
non-connected diffeomorphisms which give rise to a non-trivial actuation of 
the modular group $\hbox{Diff}({\cal M})/\hbox{Diff}_0({\cal M})$. 
This and others aspects of diffeomorphisms will not  
be further developed here but rather in a separate study. 

For any abelian group $G$, 
there is a natural group structure in the set of all $A$: 

\be (A'*A)(\gamma)=A'(\gamma)A(\gamma)\ee 
This, of course, is just another expression for the composition law
for $A$ in eq. (\ref{tGACS}). 

Now, the equation of motion $F=0$ implies that any  
$A$ can be considered a function 
on the homotopy classes $\{[\gamma]\}=\pi_1(\Re\times\Sigma)$. 
Since any loop on $\Re\times\Sigma$ 
can be continuously projected onto
$\Sigma$, we have $\pi_1(\Re\times\Sigma)=\pi_1(\Sigma)$. 

Any application, and in particular any connection, 
is completely characterized by its graph.  
Thus, since any connection is required to satisfy the condition
(\ref{compcon}), it is completely characterized by the images of the
elements of a generating subgroup of $\pi_1(\Sigma)$. Therefore we
have 

\be G_{(A)}\equiv G\otimes G\otimes\ {}_{...}^{2g}\ \otimes G\ee 
where $2g$ is the cardinal of $\pi_1(\Sigma)$. 

As is well known, the fundamental group $\pi_1(\Sigma)\equiv\{[\alpha]\}$  
of any closed surface $\Sigma$ is generated by a finite-dimensional subset 
$P_\Sigma$. 
The generator subset $P_\Sigma$ can be decomposed into two
non-intersecting subsets $P, \>\bar{P}$ 
such that to any $[\alpha]\in P$ there is associated 
a unique $[\bar{\alpha}]\in\bar{P}$ (and the other way round)  
so that there exists a representative $\alpha$ of 
$[\alpha]\in P$ and a representative 
$\bar{\alpha}$ of $[\bar{\alpha}]\in\bar{P}$ which 
intersects the one with the other exactly once.   
This property gives in fact a natural Poisson 
structure to the fundamental group of orientable surfaces
[Although as far as we know   
this analysis has not been considered in the literature, 
it would be useful to study, by also considering improper loops; that is, 
loops that begin and end in the puntures, 
how much of our analysis can be extend to 
surfaces $\Sigma$ with punctures].

For the sake of clarity we shall restrict ourselves to the groups
$\Re$ and $U(1)$. 
In both case, $\Re$ or $U(1)$, any connection is identified 
with a pair of vector ${\bf a},\>{\bf \bar{a}}$  

\be {\bf a} = \left(a_1, a_2,...a_g\right),\quad 
\bar{\bf a}=\left(\bar{a}_1, \bar{a}_2,...\bar{a}_g\right)\ee
where 

\bea A([\alpha_i])&=&\e^{2\pi a_i},\qquad \hbox{if}\>\> G=\Re\nonumber\\
A([\alpha_i])&=&\e^{i2\pi a_i}, \qquad \hbox{if}\>\> G=U(1)\eea 
The numbers $a_i, \bar{a}_i$
are (local) parameterizations of the connection. 

In the non-compact case, $G=\Re$, there are no constraints. 
The quantizing group is simply 

\bea {\bf a}'' &=& {\bf a}' + {\bf a}\nonumber\\
{\bf \bar{a}}'' &=& {\bf \bar{a}}' + {\bf \bar{a}}\label{qgcs}\\
\zeta''&=&\zeta'\zeta\exp{i\Omega\left(({\bf a'},\bar{\bf a'}),
\>({\bf a},\bar{\bf a})\right)}\nonumber\eea 
with 

\be \Omega\left(({\bf a}',\bb{a}'),\>({\bf a},\bb{a})\right) = 
{\pi k}\sum_{i\in 
P}\left({\bf a}'\cdot\bb{a}-{\bf a}\cdot\bb{a}'\right)\ee 
It is merely a Heisenberg-Weyl-like group whose
quantization is straightforward. 

\subsection{Quantization of the $U(1)$ Chern-Simons model}

The quantizing group for the $U(1)$ Chern-Simons theory 
is also given by (\ref{qgcs}) with a canonical product of the form: 

\be \Omega\left(({\bf a}',\bb{a}'),\>({\bf a},\bb{a})\right) = 
-{\pi k}\sum_{i\in 
P}\left({\bf a}'\cdot\bb{a}-{\bf a}\cdot\bb{a}'\right)\label{cpcsc}\ee 

This case is more involved and more subtle due to the 
non-trivial topology of the group $U(1)$. This non-trivial topology requires, 
in the present case, that 
two numbers $a_i\quad ({\bar{a_i}})$ that differ by an 
integer $n_i\quad(\bar{n}_i)$ have to be considered as 
equivalent. The equivalence 

\be a_i\sim a_i + n_i\qquad {\bar{a}}_i\sim {\bar{a}}_i + {\bar{n}}_i, 
\quad n_i,\>\> {\bar{n}}_i \in Z\label{aiki}\ee 
should be seen as a symmetry of the theory under gauge 
transformations which are not connected to the identity. 
The commutator of two group elements is given by: 

\be [({\bf a}',\>\bb{a}',\>\zeta'),\>({\bf a},\>\bb{a},\>\zeta)]= 
\left(0,0,\exp\{-i2\pi k({\bf a}'\cdot\bb{a} -
{\bf a}\cdot\bb{a}')\}\right) \ee 

From now on, and for the sake of simplicity,  
we shall deal with a single coordinate-momentum pair  
$(a_i,\>\bar{a}_i)$ or, what is the same, we shall restrict ourselves
to one of the handles $(g=1)$ of the surface.  
The total Hilbert space ${\cal H}$ will clearly be: 

\be  {\cal H}=\otimes_{i=1,...g}{\cal H}_i\label{totalH} \ee 
where ${\cal H}_i$ is the Hilbert space associated with the $i$th  
coordinate-momentum pair ($\equiv$ handle).  

The gauge invariance (\ref{aiki}) is incorporated 
into the quantum theory by considering the constraint 
subgroup $\widetilde{T}$ to be the following one:  

\be \widetilde{T}=\{(n,\bar{n},\zeta),\quad n,\bar{n}\in Z\}  \ee 

We shall consider only the case in which   
$k$ is a rational number; $k=\frac{p}d$ with $p$ and 
$d$ relative prime integers, $d>0$. 

\v5mm
\noindent{\bf Representing the constraint subgroup $\widetilde{T}$} 
\v3mm
The ($\widetilde{T}$-)characteristic subgroup is 

\be C= \{(dn,d\bar{n},1),\>n,\bar{n}\in Z \}\label{CT}\ee
and it is easy to show that any ($\widetilde{T}$-)polarization 
subgroup $P$ can be written in the form: 

\be P\equiv P_{p/q\bar{q}}= \{(qn,\bar{q}\bar{n},1),\>n,
\bar{n}\in Z\}\label{Tpolarization}\ee
where $q, \bar{q}$ are any two natural numbers such that $q\bar{q}=d$. 

To impose the polarization conditions properly we need
to know the general representation of the polarization
subgroup. Since these (sub)groups are abelian and finitely generated,
its irreducible representations are given by: 

\be D\left((qn,\bar{q}\bar{n},1)\right)=\e^{-i2\pi \bar{r}n}\e^{i2\pi 
r\bar{n}},\qquad r\in[0,\>1),\quad \bar{r}\in[0,\>1)
\ee 

The polarization conditions are: 

\bea \Psi_{p/q\bar{q}}(a +qn,\bar{a}, \zeta\exp{\{ik\pi qn\bar{a}\}})&=& 
\e^{-i2\pi\bar{r}n}\Psi_{p/q\bar{q}}(a,\bar{a},\zeta), \>r\in[0,1)\nonumber\\ 
\Psi_{p/q\bar{q}}(a,\bar{a}+ \bar{q}\bar{n}, 
\zeta\exp{-\{ik\pi \bar{q}\bar{n}a\}})&=& 
\e^{i2\pi r\bar{n}}\Psi_{(p/q\bar{q}}(a,\bar{a},\zeta),\>r\in[0,1) \eea 

These conditions imply that there are only $q\times\bar{q}=d$ 
independent wave functions; that is, the Hilbert space has dimension $d$. 
A natural basis is given by: 

\be B_{p/q\bar{q}}= \{  |l,\bar{l}> \}_{l=0,...q-1,\>\bar{l}=0,...\bar{q}-1}
\label{basisB}\ee 
where 

\be |l,\bar{l}>(n,\bar{n},\zeta)=\zeta\delta_{l,n}\>\delta_{\bar{l},\bar{n}}, 
\quad\forall\>\> {n=0,...q-1,\>\bar{n}=0,...\bar{q}-1}. \ee 

The action of the group operators $P_{(n,\bar{n},\zeta)}$ 
in this basis is generated by the following ones: 

\bea P_{(n,0,1)}|l,\bar{l}>&=& \e^{-i\pi k n\bar{l}}|l-n\>\bar{l}>, 
\quad\forall n\in Z\nonumber\\
P_{(0,\bar{n},1)}|l,\bar{l}>&=& 
\e^{i\pi k \bar{n}{l}}|l,\bar{l}-\bar{n}>, 
\quad\forall\bar{n}\in Z\label{Tact1}\\ 
P_{(0,0,\zeta)}|l,\bar{l}>&=& 
\zeta|l,\bar{l}>, \quad \zeta\in U(1)\nonumber\eea 
where the following equivalence conditions have to be taken into account: 

\bea |l-qn,\bar{l}>&=&\e^{-i\pi k \bar{l}qn}\e^{-i2\pi \bar{r}n}|l,\bar{l}>, 
\quad \forall n\in Z\nonumber\\
|l,\bar{l}-\bar{q}\bar{n}>&=&\e^{i\pi kl\bar{q}\bar{n}}
\e^{i2\pi r\bar{n}}|l,\bar{l}>,\quad \forall 
\bar{n}\in Z\label{Tact2}\eea
\v5mm
\noindent{\bf Constraint quantization}
\v3mm

Once we know the irreducible representations of $\widetilde{T}$ 
we can carry out the (constraint) quantization of the $U(1)$ 
Chern-Simons model.  
 
Let us choose as polarization the subgroup 

\bea P=\{(a, \bar{a},1)/a=0\}\eea 

The $P$-polarized $U(1)$-functions are given by: 

\be
\Psi(a,\bar{a},\zeta)=\zeta\exp\{ik\pi a
\bar{a}\}\varphi(a)\label{polarizedB}\ee

Now we are ready to impose the contraining conditions. 
As we already know the irreducible representations of $\widetilde{T}$,  
we can straightforwardly impose the constraining conditions 
on our wave functions. However, since $\H_{|00>}$, the vacuum subspace of 
the representations of $\widetilde{T}$, is, by construction, 
invariant under the ($\widetilde{T}$-)polarization 
subgroup $P_{p/q\bar{q}}$ in eq. (\ref{Tpolarization}),  
we shall firstly consider the action of this 
subgroup on the polarized wave functions. 

Moreover, since the operators 

\be P_{(n,\bar{n},1)}, n=1,...q,\>\>\bar{n}=1,...\bar{q}\ee 
behave, in the representation of $\widetilde{T}$, as step operators, we can 
limit ourselves to the vacuum subspace of the 
$\widetilde{T}$-representation and generate, afterwards, 
the whole Hilbert space by repeated 
application of these step operators. 

Therefore, the constraining conditions, which are 
produced by the ($\widetilde{T}$-)polarization 
subgroup (\ref{Tpolarization}), together with eq. 
(\ref{Tact1},\ref{Tact2}), provide us with the full Hilbert space of  
$\widetilde{T}$-constrained wave functions. 

Thus, let us consider the action, from the left, of the 
($\widetilde{T}$-)polarization subgroup $P_{p/q\bar{q}}$ on the functions 
in the vacuum subspace of the representation of $\widetilde{T}$.  
This gives rise to the following two conditions: 

\bea \Psi(qn+a,\bar{a}, \zeta\exp{\{-ik\pi qn\bar{a}\}})&=& 
\e^{-i2\pi \bar{r}n}\Psi(a,\bar{a},\zeta), \>r\in[0,1)\nonumber\\ 
\Psi(a,\bar{q}\bar{n} +\bar{a}, 
\zeta\exp{\{ik\pi \bar{q}\bar{n}a\}})&=& 
\e^{i2\pi r\bar{n}}\Psi(a,\bar{a},\zeta),\>r\in[0,1) \eea 

The first  condition implies for polarized wave functions 

\be \varphi(a+qn)=
\e^{-i2\pi \bar{r}n}\varphi(a)\label{quasiperiodic}\ee 

The other condition implies that the wave functions $\varphi$ are 
supported only on the connections $a$ that obey  

\be \frac{p}{q}a-r\in Z\ee 
Therefore the wave functions $\varphi$ are of the form 

\be \varphi(a) 
= \sum_{s\in Z} B_s 
\delta{(\frac{p}{q}a-r-s)} \label{B1}\ee 
where the numbers $B_s$ are not arbitrary but are required to
satisfy the quasiperiodicity condition 

\be B_{s+p}=\e^{-i2\pi \bar{r}}B_s\label{B2}\ee 
Therefore, in the sum (\ref{B1}) there are only $p$ independent complex
numbers. 

Thus, the Hilbert subspace $\H_{|00>}$ has dimension $p$. 
Now if we repeatedly apply to this subspace the operators 
$P_{(n,\bar{n},1)}$, which generate the whole $T$, 
we generate a Hilbert space $\H^{r,\bar{r}}_{\frac{p}{q\bar{q}}}$ 
with finite dimension $p\times q\times \bar{q}=p\times d$. 
We have thus recovered the well known fact that compact 
phase spaces give rise to
finite-dimensional Hilbert spaces \cite{[Field]}. 

The good operators split naturally into 
two subgroups: firstly, the subgroup 
$\widetilde{B}_{|00>}$ which is made with the operators 
that preserve the subspace $\H_{|00>}$, and, secondly, the 
subgroup $\widetilde{T}$ which transforms 
the subspace $\H_{|00>}$ into the subspaces 
$\H_{|l,\bar{l}>}$. 

It is easy to show that the subgroup $\widetilde{B}_{|00>}$ 
is the maximal subgroup of $\tG$ which obeys  

\be Ad(\tG)[P_{\widetilde{T}},
\>\widetilde{B}_{|00>}]\subset P\label{buenos1}\ee 
In the particular case at hand 
this condition reduces to  

\be [P_{\widetilde{T}},\>\widetilde{B}_{|00>}]
=\{1_{\tG}\}\label{buenos2}\ee 
and implies 

\be \widetilde{B}_{|00>}=\{(\frac{q}pn,\frac{\bar{q}}p\bar{n},
\zeta)/\>n,\bar{n}\in Z\}\label{buenos}\ee 
Therefore, the subgroup $\widetilde{B}$ 
of good operators is given by 

\bea \widetilde{B}&\equiv& \widetilde{B}_{|00>} + \widetilde{T}\nonumber\\
&=&\{(\frac{q}pn,\frac{\bar{q}}p\bar{n},
\zeta)/\>n,\bar{n}\in Z\}\oplus\{(m,\bar{m},\zeta)/\>m,\bar{m}\in Z\}\\
&=& \{(\frac{n}p,\frac{\bar{n}}p,\zeta)/\>n,\bar{n}\in Z\}\nonumber\eea 

Therefore, imposing the condition that the Hilbert space must be in   
a single irreducible representation of $\widetilde{T}$ forces us  
to only represent a subgroup $\widetilde{B}$ 
(in the present case, discrete) 
of the whole $\tG$. Applying to this Hilbert space 
operators which are not in $\widetilde{B}$  
will produce states in different representations of $\widetilde{T}$. 

The operators which are not in $\widetilde{B}$ can be classified as 

\be P_{(s',\bar{s'},\zeta)}\quad \hbox{with}\>s',\bar{s'}\in (0,\frac1p)\ee

Now, it is easy to show that 

\be P_{(s',\bar{s'},1)}\H^{r,\bar{r}}_{\frac{p}{q\bar{q}}}=
\H^{r+\frac{p}{q}s',\bar{r}+\frac{p}{\bar{q}}\bar{s'}}_{\frac{p}{q\bar{q}}}\ee 

Therefore, the Hilbert space $\H_{\frac{p}{q\bar{q}}}$ which represents 
the whole $\tG$ splits into a (continuum) sum 

\be \H= \oplus_{s,\bar{s}}
\H^{r+s,\bar{r}+\bar{s}}_{\frac{p}{q\bar{q}}},
\quad s\in(0,\frac1q),\>\bar{s}\in(0,\frac1{\bar{q}})\ee  

Finally, there is a noteworthy point to be discussed. 
The approach to the 
quantum theory in the present subsection has led us  
to an irreducible representation $D^{r,\bar{r}}_{\frac{p}{q\bar{q}}}$ 
of a subgroup of good operators $\widetilde{B}$. 
Instead, we could have determined 
this subgroup $\widetilde{B}$ firstly, 
and have quantized it afterwards (by applying 
the algebraic GAQ formalism). 
It is interesting to point out that in 
this way we would have obtained representations of $\widetilde{B}$ 
which would be different from the ones we have actually obtained. 
These representations can arise, for instance, 
by taking as $\widetilde{T}$-polarization 
$P_{\frac{p}{q\bar{q}}}= \{(qn,\bar{q}\bar{n},1)/n,\bar{n}\in Z\}$ 
and as polarization 

\be P^{u,\bar{u}}_{\frac{p}{q',\bar{q'}}}= 
\{(q'\frac{n}u,\bar{q'}\frac{\bar{n}}{\bar{u}},1)/n,\bar{n}\in Z\}\ee 
where $u\in N,\bar{u}\in Z/\>u\bar{u}=p$,  
$q',\bar{q'}\in N/\>q'\bar{q'}=d$ and, in general, $q'\> (\bar{q'})$ 
might be taken to be different from $q\> (\bar{q})$ (the representations 
we have found in the present subsection are the ones with $u=p,\bar{u}=1$ 
and $q'=q\>, \bar{q'}=\bar{q}$). This way of proceeding 
would constitute a refined version 
of the approaches in which the 
constraints are imposed firstly and the quantization is carried out 
afterwards. 

\section{Final comments. Perspectives}

We have further developed the algebraic and configuration-space  
pictures of the GAQ formalism of group quantization. 
We have combined both in order to make a comprehensive
and completely general analysis of the theory of linear fields. 
We have shown that, for linear fields, the formalism 
is extremely poweful and this power is best employed 
when the pictures just mentioned are combined. 
It has also been shown that the formalism is specially 
well suited to deal with topological issues 
(in this respect see also \cite{[frachall]}). 

We would like to remark here that the GAQ formalism can, 
in principle, be applied to any group. It gives as a result a
quantum dynamical system. However, for an arbitrary group, it is
unclear what physical interpretation, if any, the resulting dynamical
system will have. On the other hand, classical systems with a clear 
physical interpretation are commonly described, 
not by a group, but by a Lagrangian or a set
of differential equations. How to go from Lagrangian ($\sim$
differential equations) to a quantizing group 
(and the other way round) is an important question in 
the GAQ formalism but not much is known yet about its general answer. 
The present paper, however, addresses this question for the
case of linear fields. It turns out that for linear fields the set of 
solutions of the equations of motion -- that is, the
(covariant) phase space of the theory \cite{[cps],[reduced]} --, when 
extended, is a suitable quantizing group. 

A particularly attractive direction of development 
is, therefore, towards non-linear fields. 
However, there appear to be obstructions 
for the phase space of non-linear
fields to have a group structure. In particular,   
ref. \cite{[config]} presented 
indications that for non-abelian current
groups with group law of a pointwise type, any 
equation of motion which preserves the group structure would 
have to be first order in derivatives of the space-time co-ordinates. 
A rigorous theorem is, however, still 
lacking and, after all, first-order equations may give plenty of room for
interesting developments as recent studies, 
relevant to our approach, indicate \cite{[Geroch]}. 
On the other hand, constraint quantization 
might be used to circumvent the problem of not 
having a group structure in the phase space of the theory.  
In addition to all this, it was also 
shown in ref. \cite{[config]} that for
some current groups with group laws of a non-pointwise type,  
we can actually find higher-order differential 
equations which preserve them. 

Let us finally consider the case of non-linear gauge fields. 
For linear gauge fields,  if $A,\>A':\L_\M \rightarrow G$
are connection and we define a composition law $*$ 
by means of the equality 

\be (A'*A)(\gamma)=A'(\gamma)A(\gamma),\label{cs20}\ee 
then $A'*A$ is also a connection. As we have shown the composition 
law $*$ is also compatible with the equations of motion, and thus defines 
the natural group law for the theory. 
However, when $G$ is non-abelian,  $A''=A'*A$ defined by eq. (\ref{cs20}) 
does not satisfy the condition 

\be A''(\gamma'\circ\gamma)=A''(\gamma')A''(\gamma)\ee 
and thus $A''$ is not a connection. Therefore, 
a ``naive'' extension of the configuration-space approach to 
non-abelian gauge fields is problematic even before
the equations of motion are considered. 

Summarizing we would say that, because of obstructions which arise,   
the analysis we have performed in this paper for linear fields cannot be 
straightforwardly extended to non-linear fields. However, 
the real importance of the obstructions is still not clear and 
further investigations are in order.\hfil\break 

\noindent {\bf Acknowledgements:} M. N. is grateful to the
Imperial College, where this paper has mainly been written, 
for its hospitality. M.N. is also grateful to the Spanish MEC, CSIC and 
IMAFF (Madrid) for a research contract. M.C. is grateful 
to the Spanish MEC for a FPI fellowship.

\end{document}